\documentclass{article}
\usepackage{spconf,amsmath,graphicx}
\usepackage{cite}
\usepackage{threeparttable}
\usepackage{amsfonts}
\usepackage{booktabs}
\usepackage{url}
\usepackage{xcolor}
\usepackage{multirow}
\usepackage{makecell}
\usepackage{subfig}
\usepackage{float}
\usepackage[colorlinks,linkcolor=red]{hyperref}
\title{Time-weighted Frequency Domain Audio Representation with GMM Estimator for  Anomalous Sound Detection}
\name{Jian Guan$^{1*}$\thanks{*Corresponding Author}, Youde Liu$^2$,  Qiaoxi Zhu$^3$, Tieran Zheng$^2$, Jiqing Han$^2$, Wenwu Wang$^4$
\thanks{This work was partly supported by the Natural Science Foundation of Heilongjiang Province under Grant No. YQ2020F010, 
and a Newton Institutional Links Award from the British Council with Grant No. 623805725. For the purpose of open access, the authors have
applied a creative commons attribution (CC BY) licence to any author
accepted manuscript version arising.
}}
\address{
  $^1$Group of Intelligent Signal Processing, College of Computer Science and Technology, \\Harbin Engineering University,  China\\
  $^2$School of Computer Science and Technology, Harbin Institute of Technology, China\\
  $^3$Centre for Audio, Acoustics and Vibration, University of Technology Sydney,  Australia\\
  $^4$Centre for Vision Speech and Signal Processing, University of Surrey,  UK}

\begin{document}
%
\maketitle
\begin{abstract}
%
Although deep learning is the mainstream method in unsupervised anomalous sound detection, Gaussian Mixture Model (GMM) with statistical audio frequency representation as input can achieve comparable results with much lower model complexity and fewer parameters. Existing statistical frequency representations, e.g. the log-Mel spectrogram's average or maximum over time, do not always work well for different machines. This paper presents Time-Weighted Frequency Domain Representation (TWFR) with the GMM method (TWFR-GMM) for anomalous sound detection. The TWFR is a  generalized statistical frequency domain representation that can  adapt to different machine types, using the global weighted ranking pooling over time-domain. This allows GMM estimator to recognize anomalies, even under domain-shift conditions, as visualized with a Mahalanobis distance-based metric. Experiments on DCASE 2022 Challenge Task2 dataset show that our method has better detection performance than recent deep learning methods. TWFR-GMM is the core of our submission that achieved the 3rd place in DCASE 2022 Challenge Task2.
\end{abstract}
\section{Introduction}
\label{sec:intro}
%
Anomalous sound detection (ASD) identifies whether a target machine is anomalous from its emitted sound. In unsupervised ASD, the detector is often trained with only normal sounds, due to the fact that anomalous sounds can be rare and diverse, and thus hard to capture in practice. Unsupervised ASD for machine condition monitoring has been Task2 of the Detection and Classification of Acoustic Scenes and Events (DCASE) Challenge since 2020 \cite{Koizumi_DCASE2020_01}. Different from Task2 of DCASE 2020, Task2 of DCASE 2021 and DCASE 2022 focuses on the domain shift problem, where the acoustic characteristics of the training and test data are different \cite{Kawaguchi_arXiv2021_01, Dohi_arXiv2022_02}.

Deep learning methods are the mainstream methods for anomalous sound detection. Autoencoder (AE) based method is the baseline in DCASE Task2 for unsupervised ASD~\cite{Koizumi_DCASE2020_01,Kawaguchi_arXiv2021_01,Dohi_arXiv2022_02}. It uses the AE model to learn the distribution of normal sounds by reconstructing the log-Mel spectrograms of normal sounds and using the reconstruction errors to derive anomaly scores. This method is then improved by incorporating device information to learn better feature representation of normal sounds \cite{Giri2020a, liu2022anomalous}. For example, MobileNetV2 \cite{sandler2018mobilenetv2} uses a self-supervised machine identity (ID) classifier to learn a better representation of normal sounds, and uses the negative log-likelihood (NLL) of the corresponding ID as the anomaly score \cite{Giri2020a}. 
Instead of calculating the anomaly score from the output probability of ID classification, another method, i.e., ResNet-GMM \cite{wilkinghoff2021sub} uses the Gaussian Mixture Model (GMM) as the anomaly estimator, following a self-supervised ID classifier constructed by ResNet \cite{he2016deep}. Here, GMM is used to cluster the audio features of normal sound by fitting the distribution of normal sounds as a mixture of a finite number of Gaussian distributions. 

\begin{figure*}[!ht]
    \centerline{
    \includegraphics[width=.92\textwidth]{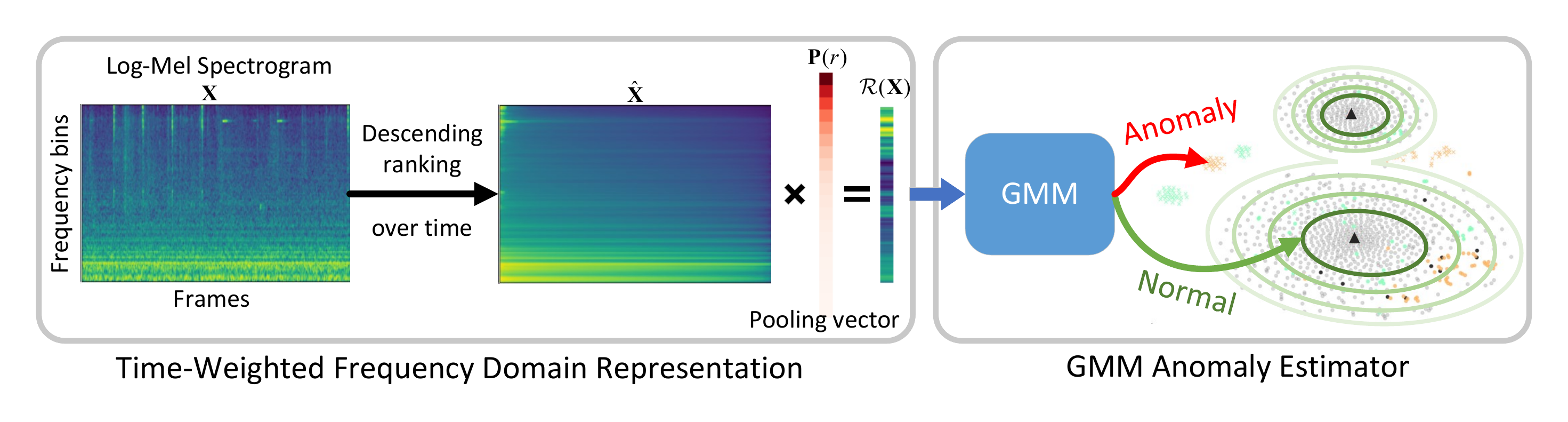}}
    \vspace{-2.5mm}
    \caption{Proposed Time-Weighted Frequency Domain Representation with Gaussian Mixture Model (TWFR-GMM) method. TWFR is extracted from descending ranking the log-Mel spectrogram at time frames per frequency band and multiplying with a weighting vector from global weighted ranking pooling $\mathbf{P}(r)$ to form a dimension-reduced audio representation $\mathcal{R}(\mathbf{X})$.}
    \vspace{-3mm}
    \label{fig:1}
\end{figure*}
%
%
Non-deep learning method for ASD was initially attempted by using the GMM estimator with inputs from the average or maximum pooling on the log-Mel spectrogram over time dimension \cite{wilkinghoff2021sub}, namely Mean-GMM and Max-GMM. Their inputs are frequency domain audio representations considering simple statistical features (mean or maximum) over time, rather than extracted from a neural network, e.g. ResNet. These models often involve fewer parameters and lower complexity than deep learning models \cite{suefusa2020anomalous, Giri2020a, dohi2021flow}, but offer a comparable detection performance.
However, the simple average or max pooling representation does not always perform well on different machine types. For example, Mean-GMM performs better than Max-GMM for ToyCar, Fan, Gearbox, while Max-GMM performs much better than Mean-GMM for ToyTrain and Valve, when tested on the DCASE 2022 Task2 dataset.

Due to the varying time-frequency distribution of audio amongst different machines, it is desirable to develop an adaptive statistical model for time-frequency representation of audio signals to suit different machines. In addition, it is desirable to maintain the advantage of low model complexity over the deep learning approaches, while applying proper trade-offs between the stationary and non-stationary parts of the audio signals.

As a solution, we propose a Time-Weighted Frequency Domain Representation 
 (TWFR) with Gaussian Mixture Mode (TWFR-GMM) method for unsupervised ASD in this paper. The TWFR is extracted from global weighted ranking pooling (GWRP) \cite{kolesnikov2016seed} on log-Mel spectrogram over time dimension, which considers the whole time range and applies more weight to the time frames with higher sound energy. To investigate the effectiveness of the proposed method, we design a Mahalanobis Distance based metric for t-distributed stochastic neighbour embedding (t-SNE) \cite{van2008visualizing} visualization of TWFR-GMM. The t-SNE visualization shows the clustering effect of GMM on the training and test sounds. It verifies that the TWFR directly derived from the normal machine sound can be used to distinguish anomalies well, even under domain-shift conditions. Experiments on DCASE 2022 Challenge Task2 dataset show that our method has better detection performance than state-of-the-art deep learning methods. TWFR-GMM is the core of our submission that achieved the top 3 in DCASE 2022 Challenge Task2. We release the source code for the reproducibility of our work at \url{https://github.com/liuyoude/TWFR-GMM}.
\begin{table*}[!htbp]
\vspace{-2mm}
    \setlength{\abovecaptionskip}{0cm}
    \centering
    \caption{Performance comparison in terms of AUC (\%) and pAUC (\%) on the development dataset of DCASE 2022 Task2. Average: the average of the AUC and pAUC values for all machine types. }
    \resizebox{\textwidth}{!}{
    \begin{tabular}{lcccccccccccccccc}
        \toprule
        \multirow{2}{*}{Methods} & \multicolumn{2}{c}{ToyCar} & \multicolumn{2}{c}{ToyTrain} & \multicolumn{2}{c}{Fan} & \multicolumn{2}{c}{Gearbox} & \multicolumn{2}{c}{Bearing} & \multicolumn{2}{c}{Slider} & \multicolumn{2}{c}{Valve} & \multicolumn{2}{c}{Average}\\
        \cmidrule(lr){2-3} \cmidrule(lr){4-5} \cmidrule(lr){6-7} \cmidrule(lr){8-9} \cmidrule(lr){10-11} \cmidrule(lr){12-13} \cmidrule(lr){14-15} 
        \cmidrule(lr){16-17} 
         & AUC & pAUC & AUC & pAUC & AUC & pAUC & AUC & pAUC & AUC & pAUC & AUC & pAUC & AUC & pAUC & AUC & pAUC \\
        \midrule

        \multicolumn{3}{l}{\textit{\textbf{Deep Learning} based methods:}\vspace{2mm}}\\


        MobileNetV2 \cite{Dohi_arXiv2022_02} & 
        55.54 & 52.27 & 51.57 & 51.51 & 59.48 & 56.89 & 62.70 & 56.03 & 60.25 & 57.14 & 51.69 & 54.67 & 62.14 & 62.41 & 57.62 & 55.85 \\
        
        AE  \cite{Dohi_arXiv2022_02} & 
        62.61 & 52.74 & 49.83 & 50.48 & 62.89 & 57.52 & 65.78 & 58.49 & 56.40 & 51.98 & 62.81 & 55.78 & 50.73 & 50.36 & 58.72 & 53.91\\



        STgram-MFN \cite{liu2022anomalous} & 
        48.62 & 49.91 & 52.10 & 49.62 & 64.42 & 60.95 & 75.97 & 64.21 & \textbf{78.58} & \textbf{64.28} & 79.46 & 63.22 & 70.80 & 63.32 & 67.14 & 59.36 \\







        \midrule
        \multicolumn{3}{l}{\textit{\textbf{Non-Deep Learning} based methods:}\vspace{2mm}}
        \\

        Mean-GMM \cite{wilkinghoff2021sub} & 
        79.33 & 58.93 & 58.74 & 52.74 & \textbf{71.44} & \textbf{62.13} & 79.62 & 65.37 & 69.41 & 54.74 & 78.57 & 63.62 & 52.39 & 50.54 & 69.93 & 58.30 \\

        Max-GMM \cite{wilkinghoff2021sub} & 
        60.21 & 51.19 & 67.34 & 55.28 & 62.51 & 52.71 & 71.40 & 54.68 & 64.15 & 52.29 & 82.65 & 66.49 & 91.87 & 69.86 & 71.45 & 57.50 \\

        \textbf{TWFR-GMM}  & 
        80.65 & 58.12 & 67.81 & 59.07 & \textbf{71.44} & \textbf{62.13} & 80.23 & 64.69 & 69.41 & 54.74 & 87.96 & 73.33 & \textbf{92.61} & \textbf{70.23} & 78.59 & 63.19\\

        \textbf{SMOTE-TWFR-GMM}  & 
        \textbf{84.70} & \textbf{60.55} & \textbf{68.02} & \textbf{59.69} & \textbf{71.44} & \textbf{62.13} & \textbf{82.22} & \textbf{65.85} & 70.80 & 54.22 & \textbf{89.91} & \textbf{76.74} & \textbf{92.61} & \textbf{70.23} &
        \textbf{79.96}& \textbf{64.20} \\

        \bottomrule
        \bottomrule
    \end{tabular}
    }
    \label{tab:4}
    \vspace{-2mm}
\end{table*}
%


%
\section{Proposed Method}
\label{sec:method}
The proposed TWFR-GMM method for unsupervised ASD is illustrated in Figure \ref{fig:1}, with details of the audio representation module, Time-Weighted Frequency Domain Representation (TWFR) in Section \ref{subsec:twfr} and GMM estimator in Section  \ref{subsec:gmm-estimator}.
%
%
\subsection{Time-Weighted Frequency Domain Representation}
\label{subsec:twfr}
%
%
Existing statistical frequency representation is simply the average or maximum of the log-Mel spectrogram over time, and not always work
well for different machines. Specifically, Max-GMM only considers the time frame with the maximum sound energy, while ignoring all the other time frames. As a result, the stationary feature, one of the critical features of normal machine sounds, is not counted. On the contrary, mean-GMM evenly accounts for every time frame over the whole time span, which captures the stationary feature well, but may not be able to capture transient features for short-term signals, due to the averaging operation. However, the transient features can differ significantly between the normal and abnormal sound.

The proposed TWFR-GMM maintains the advantage in low model complexity over the deep learning approaches by applying simple time-domain weighting pre-selected to adapt to each machine type. This weighting is used for trading off the stationary and non-stationary audio signals to form a simple statistical frequency domain representation of the audio signal. In detail, we apply the global weighted ranking pooling (GWRP) \cite{kolesnikov2016seed} on the log-Mel spectrogram over time dimension to give more weight to the time frames with higher sound energy. Consider the log-Mel spectrogram $\mathbf{X} \in \mathbb{R}^{M \times N}$ of an audio signal with $M$ Mel frequency bins and $N$ time frames. We sort elements in its $i$-th row vector $\mathbf{X}_i \in \mathbb{R}^{1 \times N} (i = 1, 2, \cdots, M)$  (i.e., $i$-th frequency bin over time frames)
by descending ranking as $\mathbf{\hat{X}}_i$. 
It re-arranges the values over time frames at each frequency bin by energy-decreasing sequence and ignores the time sequence. The resulting sequence $\mathbf{\hat{X}} = [\mathbf{\hat{X}}_1, \mathbf{\hat{X}}_2, ..., \mathbf{\hat{X}}_M]$ is derived from $\mathbf{X}$ in a descending order of time. The TWFR extracted from the GWRP of $\mathbf{X}$ is
\begin{equation}
    \mathcal{R}(\mathbf{X}) = \mathbf{\hat{X}}\mathbf{P}(r),
    \label{eq:1}
\end{equation}
where $\mathcal{R}(\mathbf{X}) \in \mathbb{R}^{M}$ and the pooling vector
\begin{equation}
\mathbf{P}(r)=[\frac{r^0}{z(r)}, \frac{r^1}{z(r)}, ..., \frac{r^{N-1}}{z(r)}]^\text{T},
\end{equation}
with $z(r)=\sum_{n=1}^N r^{n-1}$ for normalization. The superscript T denotes the transpose operation.
%
%

The average pooling and maximum pooling are two special cases of GWRP. When $r = 0$, GWRP degenerates to maximum pooling, and when $r = 1$, GWRP becomes average pooling. Our method finds a suitable $r$ from 0 to 1 for the best detection performance of each machine type in the training stage. As a result, TWFR can be adapted for different machine types to achieve more robust audio feature representation for anomalous sound detection.
%
%
%
%
%
\subsection{Gaussian Mixture Model (GMM) Estimator}
\label{subsec:gmm-estimator}
%
%
\textbf{\textit{TWFR-GMM:}} 
GMM is used to fit the distribution of normal sounds as a mixture of a finite number of Gaussian distributions. The GMM is trained on normal sounds and detects anomalies for test sounds in terms of the negative log-likelihood defined as  
\begin{equation}
    \mathcal{A}(\mathbf{\bar{X}}) = - \max_{k\in[1,K]} log \mathcal{N}(\mathcal{R}(\mathbf{\bar{X}})|\mathbf{\mu}_{k},\mathbf{\Sigma}_{k}),
\end{equation}
where $\mathcal{N}(\mathcal{R}(\bar{\mathbf{X}})|\mathbf{\mu}_{k},\mathbf{\Sigma}_{k})$ is the $k$-th Gaussian distributions of the trained GMM, with mean vector $\mathbf{\mu}_{k} \in \mathbb{R}^{M}$ and covariance matrix $\mathbf{\Sigma}_{k} \in \mathbb{R}^{M \times M}$, and  $\bar{\mathbf{X}}$ is the log-Mel spectrogram of the test sound.
%

\noindent \textbf{\textit{SMOTE-TWFR-GMM:}} For DCASE 2022 Challenge Task2 focusing on domain shift problem, we train GMM respectively for each section of each machine type, where different sections relate to different domain shift conditions. In addition, SMOTE \cite{smote} is employed to deal with sample insufficiency by over-sampling the samples in the target domain for some machine types.

%

%
\section{Experiments and Results}
\label{sec:experiment}
\subsection{Experimental Setup}
\label{subsec:setup}
\textbf{\textit{Dataset:}} We conduct experiments on the development dataset of DCASE 2022 Challenge Task2 \cite{Dohi_arXiv2022_02}, which includes seven machine types (ToyCar, ToyTrain, Fan, Gearbox, Bearing, Slider and Valve). The dataset has three sections for each machine type (Sections 00, 01 and 02), and different sections indicate different domain shift conditions. Each audio file is 10 seconds with 16k Hz sampling rate, including machine sound and environmental noises. 

\noindent \textbf{\textit{Evaluation Metrics:}} 
%
%
%
Following DCASE 2022 Challenge Task2 \cite{Dohi_arXiv2022_02}, we use the area under the receiver operating characteristic (ROC) curve (AUC) and the partial-AUC (pAUC) over all the machine types, sections, and domains as the performance metrics. In addition, the number of parameters and floating point operations (Flops) are employed for model size and complexity evaluation respectively.
%

%
\noindent \textbf{\textit{Implementation:}} The log-Mel spectrogram is extracted from audio with the window size of 1024 samples with 50\% overlap and the Mel-filter with 128 banks. The GMM and t-SNE visualization are implemented by \textit{scikit-learn} library \cite{scikit-learn}, and the number of mixture components of GMM is 1 or 2 according to the machine type. The SMOTE is implemented by \textit{imbalanced-learn} library \cite{imlearn} and applied on part of machine types. The weighting parameter $r$ in Equation \eqref{eq:1} is 0.99, 0.81, 1.00, 0.99, 1.00, 0.88 and 0.45 for Toycar, ToyTrain, Fan, Gearbox, Bearing Slider and Valve, respectively.   

\subsection{Performance Comparison}
\label{subsec:performance}
Table \ref{tab:4} compares the proposed TWFR-GMM and SMOTE-TWFR-GMM with Mean-GMM, Max-GMM and deep learning based methods (AE \cite{Dohi_arXiv2022_02}, MobileNetV2 \cite{Dohi_arXiv2022_02} and STgram-MFN \cite{liu2022anomalous}) on the development dataset of DCASE 2022 Task2, where AE and MobileNetV2 are the baseline methods of Task 2.
%
%
%
%
%
TWFR-GMM significantly improved the average AUC and pAUC performance, compared with the deep learning based methods and the non-deep learning GMM methods. In addition, SMOTE-TWFR-GMM further improved the performance by using an over-sampling strategy in the target domain, demonstrating the effectiveness of our proposed method for ASD under domain shift. 
%
The proposed SMOTE-TWFR-GMM in this paper forms the core of the submission to the DCASE 2022 Challenge Task 2, and ranked the 3rd place in the competition \cite{GuanHEU2022}.
\begin{figure}[!htbp]
    \centerline{
    \includegraphics[width=.80\linewidth]{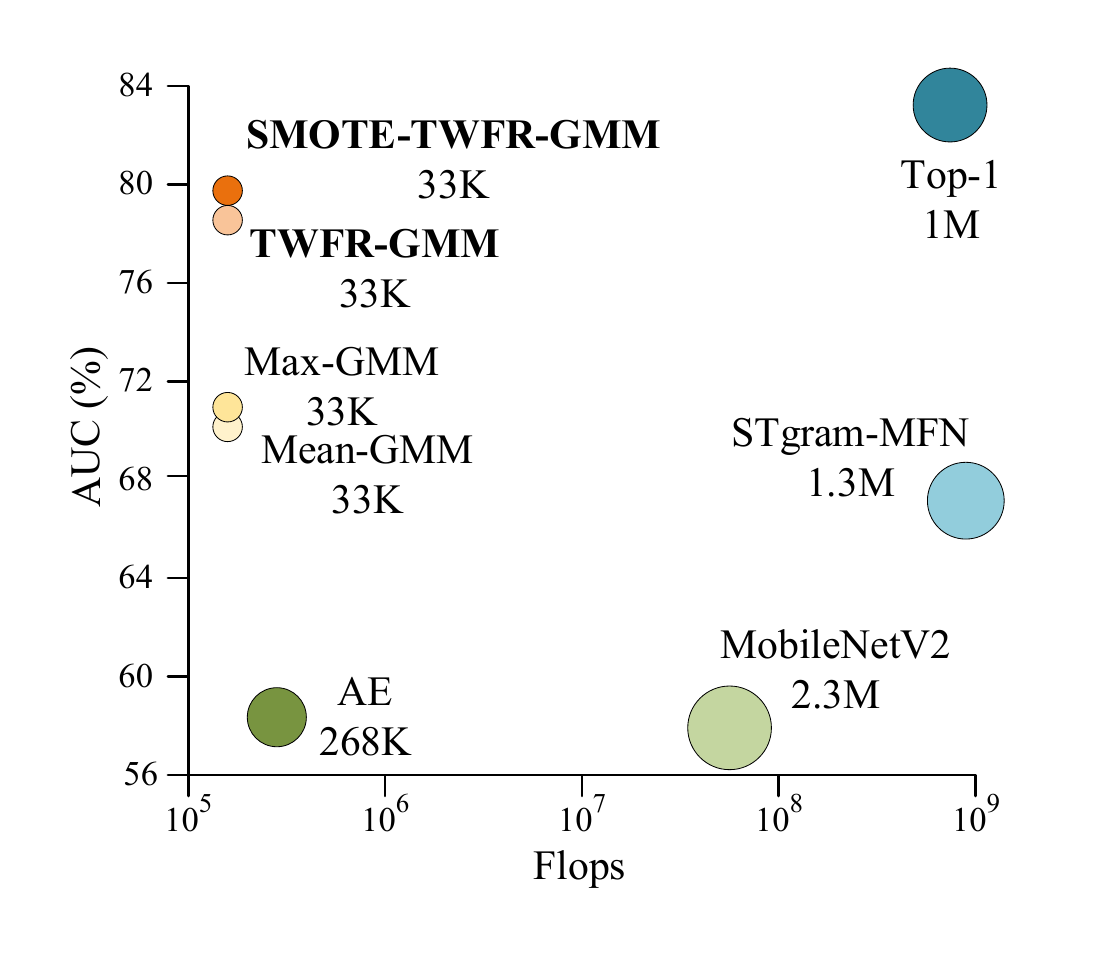}}
    \vspace{-5mm}
    \caption{
    Comparison of the model size and complexity (number of parameters and Flops), where the model size is illustrated by the size of the circles, and the number of Flops is presented in logarithmic scale.
    }
  \vspace{-3mm}
    \label{fig:flops}
\end{figure}
%
%

%
We also compared the model size and complexity of the above methods and the Top-1 method (self-supervised deep learning method) \cite{LiuCQUPT2022} in DCASE 2022 Challenge Task2, as illustrated in Figure \ref{fig:flops}. The proposed TWFR-GMM methods significantly outperform the above deep learning based methods (AE, MobileNetV2 and STgram-MFN) with much smaller model size and complexity. Compared with the Top-1 method (83.28\%), our SMOTE-TWFR-GMM (79.96\%) performs slightly worse. Our method has a much smaller model size with significantly reduced model complexity. Specifically, the Top-1 method has a model size of 4.1 M and Flops of 700 M, while our proposed method has a model size of only 33 k and Flops of only 164 k. 
%
%
This demonstrates that our proposed methods achieve better or comparable detection performance as compared with the baseline methods, with much-reduced model size and complexity.

%
%
%
\begin{figure}[!ht]
    \centerline{
    \includegraphics[width=.90\linewidth]{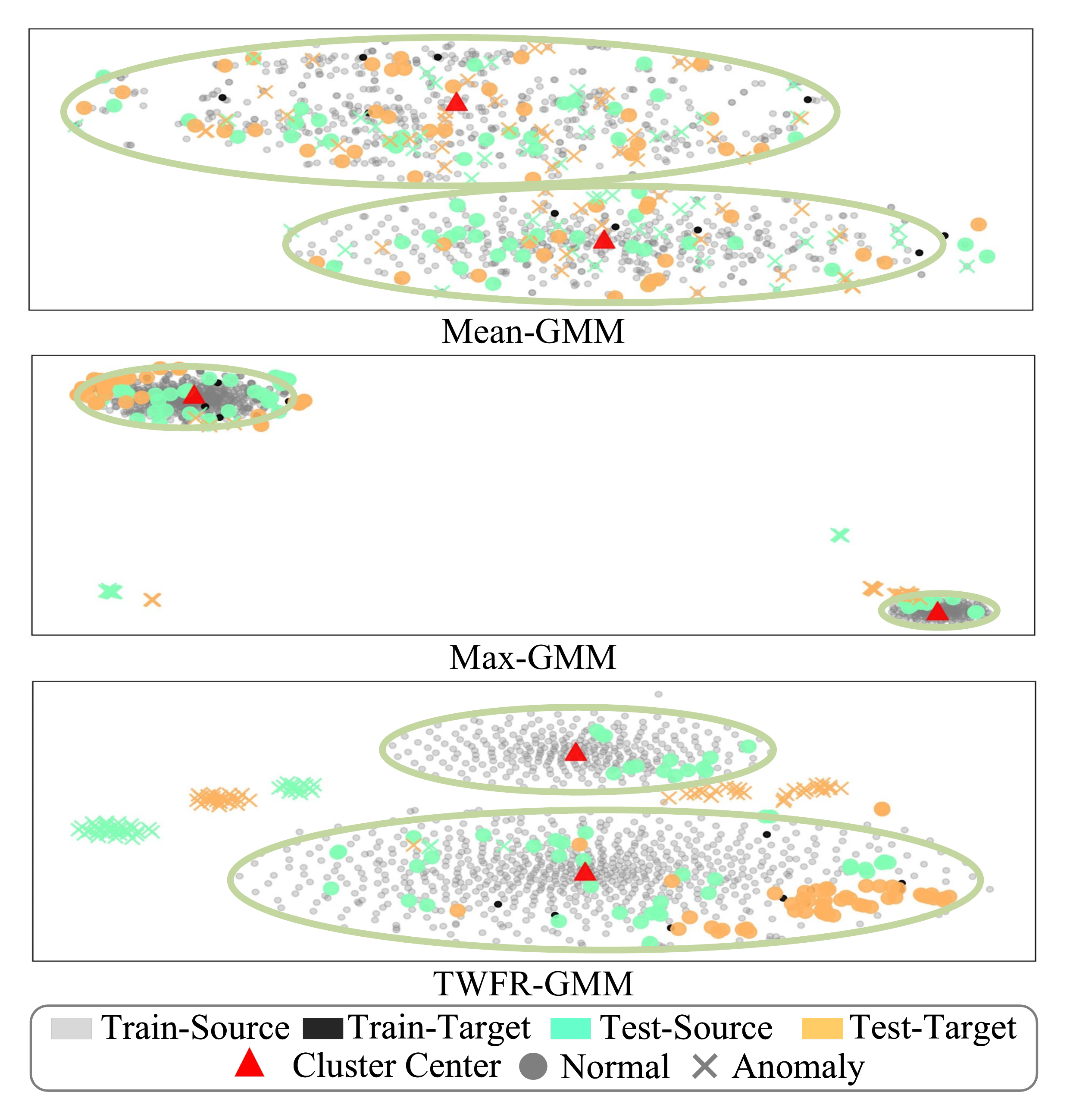}}
    \caption{The t-SNE visualization with Mahalanobis Distance based metric function of the Mean-GMM, Max-GMM and TWFR-GMM for Section 02 of Valve. Here, the mean vector of the trained GMM is adopted as the  cluster center, and  ``$\bullet$"  and ``$\times$" denote normal and anomalous samples, respectively.}
   \vspace{-2mm}
    \label{fig:tsne}
\end{figure}
%

%
\subsection{Mahalanobis Distance based Visualization Analysis}
\label{sec:visual_analysis}
%
We analyze the effectiveness of the proposed TWFR-GMM by performing visual analysis using t-distributed stochastic neighbour embedding (t-SNE) \cite{van2008visualizing} with a Mahalanobis Distance (MD) based metric function, defined as follows
%
%
%
\begin{equation}
\begin{split}
    \mathcal{M}(\mathbf{y}_1, \mathbf{y}_2) = \min_{k\in[1,K]} \sqrt{(\mathbf{y}_1 - \mathbf{y}_2)^\text{T} \mathbf{\Sigma}_k^{-1} (\mathbf{y}_1 - \mathbf{y}_2)},
\end{split} 
\label{eq:4}
\end{equation}
where $\mathbf{\Sigma}_k$ is the covariance matrix of the $k$-th mixture component of the trained GMM, and $\mathbf{y}_1$ and $\mathbf{y}_2$ are two feature vectors from both the training and test sets including source and target domain. Note that, we use the MD metric function, instead of the Euclidean Distances (ED), as the ED does not perform well, in particular, we observed that the anomalous and normal features cannot be visually separated even though the AUC is high. 

%
%

Figure \ref{fig:tsne} is the t-SNE visualization of TWFR-GMM, Mean-GMM and Max-GMM for section 02 of Valve in the development dataset of DCASE 2022 Task2, with the MD metric. Here, the distribution of the normal training sound from both source domain (Train-Source) and target domain (Train-Target) is provided to show how the detected sound from source (Test-Source) and target domain (Test-target) fit the distribution of normal sound using different methods.

In this case (i.e. section 02 of Valve), Mean-GMM cannot distinguish the normal and anomalous features, whereas Max-GMM can. However, in Max-GMM, some anomalous features are very close to normal features (i.e. just outside the clusters). In contrast, the proposed TWFR-GMM can distinguish the anomalous sound well in the MD space, despite the domain shift and the training data being sparse in the target domain.

\section{Conclusion}
\label{sec:conclusion}
%
%
This paper has presented a new method for unsupervised ASD, namely, Time-Weighted Frequency Domain Representation with Gaussian Mixture Model (TWFR-GMM). The proposed method is of low complexity, which is appealing for applications with limited computing resources. Experiments on DCASE 2022 Task2 dataset shows that the proposed method performs better than several compared methods. Furthermore, its effectiveness is analyzed by t-SNE visualization with Mahalanobis distance based metric function. Although the proposed method did not achieve the best performance on DCASE 2022 Task2, 
it provides a promising way for model refinement via improved audio representation. 


\bibliographystyle{IEEEtran}
\bibliography{strings}

\end{document}